\documentclass[a4paper,11pt]{article}
\usepackage{pos}
\usepackage{tikz}
\usetikzlibrary{positioning}
\usepackage{graphicx}
\graphicspath{{./figs/}}

\pgfdeclarelayer{bg}
\pgfsetlayers{bg,main}

\title{Flow-based sampling for lattice field theories}

\author{Gurtej Kanwar}

\affiliation{Albert Einstein Center, Institute for Theoretical Physics,\\
University of Bern, 3012 Bern, Switzerland}

\emailAdd{kanwar@itp.unibe.ch}

\abstract{Critical slowing down and topological freezing severely hinder Monte Carlo sampling of lattice field theories as the continuum limit is approached. Recently, significant progress has been made in applying a class of generative machine learning models, known as ``flow-based'' samplers, to combat these issues. These generative samplers also enable promising practical improvements in Monte Carlo sampling, such as fully parallelized configuration generation. These proceedings review the progress towards this goal and future prospects of the method.}

\FullConference{The 40th International Symposium on Lattice Field Theory (Lattice 2023)\\
July 31st - August 4th, 2023\\
Fermi National Accelerator Laboratory\\}

\tableofcontents

\begin{document}
\maketitle

\section{Introduction}
Lattice simulations play a vital role in the study of quantum field theories, allowing properties of a theory to be non-perturbatively computed from first principles. For example, lattice field theory calculations are key to understanding the Quantum Chromodynamics (QCD) sector of the Standard Model. Lattice field theory calculations require evaluating the lattice-regularized path integral,
\begin{equation}
    \left< \mathcal{O} \right> \equiv \frac{1}{Z} \int \mathcal{D}[U] \, \mathcal{O}[U] \, e^{-S[U]}, \qquad {Z} \equiv \int \mathcal{D}[U] \,  e^{-S[U]} ,
\end{equation}
where $\mathcal{O}$ is an operator of interest, $S$ is the discretized action of the theory, $\int \mathcal{D}[U]$ indicates integration over all lattice field configurations, and $Z$ is the partition function.
Almost all calculations achieve this through Monte Carlo sampling according to the distribution
\begin{equation}
    p[U] = e^{-S[U]} / Z
\end{equation}
which allows unbiased estimates of $\left< \mathcal{O} \right>$ with systematically improvable uncertainties. Traditional approaches based on Markov Chain Monte Carlo have achieved incredible progress through the use of high-performance computing up to the exascale, but the precision of many results is limited by Monte Carlo statistics. Increasing these statistics is particularly challenging at small lattice spacings where critical slowing down causes Markov Chain autocorrelations to diverge as the continuum limit is taken~\cite{Wolff:1989wq}. Furthermore, for state-of-the-art calculations, practical challenges associated with storage costs and efficient parallelization are encountered as the number of lattice sites increases~\cite{Joo:2019byq,Boyle:2022ncb}.

Recently, an approach to Monte Carlo sampling using \emph{flow-based models} has been proposed and evaluated in several lattice field theory contexts, with promising early results. Such models are based on \emph{normalizing flows}~\cite{tabak2010density,tabak2013family}, which in this context describe efficient transformations between distributions over lattice field configurations, often using machine learning techniques to parameterize specific components of transformations. Models can be constructed to use normalizing flows for a variety of tasks:
\begin{itemize}
    \item Perform efficient Monte Carlo sampling
    \item Estimate the partition function $Z$
    \item Draw correlated samples at different bare physical parameters
    \item Efficiently perform physical parameter scans
\end{itemize}
These proceedings discuss the progress in this developing field and its future prospects.

\section{Flow-based sampling}

\subsection{Normalizing flows}

\begin{figure}
    \centering
    \begin{tikzpicture}
    \node (base) at (0,0)
    {\includegraphics[height=3cm]{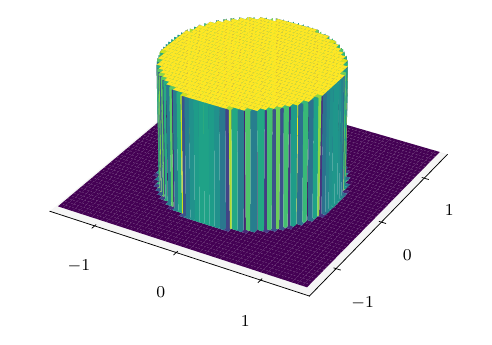}};
    \node (output) at (6,0)
    {\includegraphics[height=3cm]{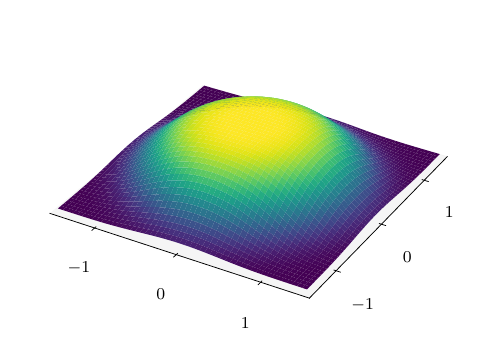}};
    \node[below=0cm of base, text width=6cm, align=center]
    {\small (Simple) Prior density\\$r(x,y) = \mathrm{Unif}_{D}(x,y)$};
    \node[below=0cm of output, text width=6cm, align=center]
    {\small (More complex) Output density\\$q(x',y') = \tfrac{1}{2\pi} e^{-(x'^2 + y'^2)/2}$};
    \draw[thick,->] (base) -- (output) node [midway,above] {\small Flow $f$};
    \end{tikzpicture}
    \caption{Example of the Box-Muller transformation, which can be considered a simple normalizing flow.}
    \label{fig:box-muller}
\end{figure}
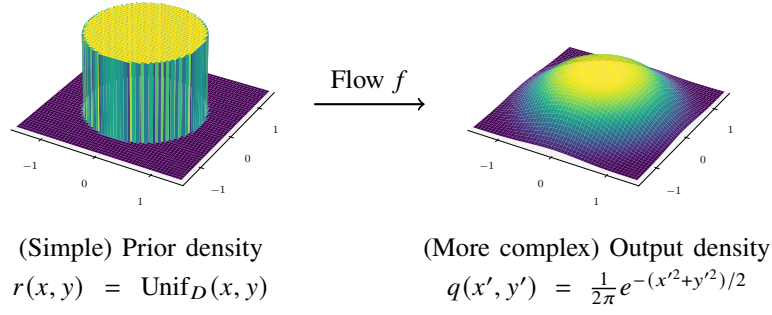

To get a feel for normalizing flows, we can consider a simple example in two variables. Starting from points $(x,y)$ distributed uniformly in the unit disk, $D = \{ x,y \, | \, x^2 + y^2 < 1 \}$, consider applying the Box-Muller transformation $(x',y') = f(x,y)$ defined in terms of $r \equiv \sqrt{x^2 + y^2}$ as~\cite{box1958note}
\begin{equation}
    x' = \frac{x}{r} \sqrt{-2 \ln r^2}
    \qquad \text{and}
    \qquad
    y' = \frac{y}{r} \sqrt{-2 \ln r^2}.
\end{equation}
This is an invertible map from the unit disk to the whole 2D plane. The action of the transformation stretches the points near the edge of the disk towards infinity while concentrating the points near the center, as shown in Fig.~\ref{fig:box-muller}. The output probability density $q(x',y')$ is given in terms of the \emph{prior density} $r(x,y) = \mathrm{Unif}_{D}(x,y)$ and the Jacobian $J_f$ of the transformation as
\begin{equation}
    q(x', y') = r(x,y) |\det J_f|^{-1} = \tfrac{1}{2\pi} e^{-(x'^2 + y'^2)/2}.
\end{equation}
This is simply the unit normal distribution over two variables. The Box-Muller transformation can thus be used to draw Gaussian-distributed variables (which are useful in many contexts) by producing samples of uniform random variables on the disk (which are easy to generate) and applying $f$.

\subsection{Flows using machine learning}

Generalizing this concept, if one can find a diffeomorphism $f$ with a tractable Jacobian between a simple distribution and a more complicated distribution of interest, then this approach can be used to efficiently sample the latter distribution. In most contexts, however, either such a transformation is not known explicitly or it is expensive to apply the function or its Jacobian. In the lattice field theory context, the trivializing map~\cite{Luscher:2009eq} is one example of a flow from a gauge theory at infinite lattice spacing to one at finite lattice spacing; however, the map and its Jacobian are defined by integrating ordinary differential equations (ODEs), which was demonstrated to be too costly to provide a significant advantage in practice~\cite{Engel:2011re}.

An alternate approach is to \emph{approximate} the target distribution $p$ by optimizing the choice of flow function $f$ within a family of functions to produce an output distribution $q$ that is close to the target distribution. Working with a variational family that is explicitly defined to ensure the function and its Jacobian determinant can be efficiently evaluated restricts to computationally tractable choices~\cite{rezende2015variational}.
Families of functions can be defined by a variety of methods, but approaches based on machine learning are appealing: neural networks can efficiently parameterize large classes of functions~\cite{hornik1989multilayer}, existing frameworks  can be used to efficiently evaluate such functions, and stochastic gradient descent can be straightforwardly applied to ``train'' (optimize) the choice of function. For a comprehensive review of progress in parameterizing normalizing flows using machine learning methods, see Ref.~\cite{papamakarios2021normalizing}.

To simplify the task of ensuring the Jacobian determinant remains tractable while defining an expressive family of flows, one of two broad approaches is commonly used:

\paragraph{Discrete learnable flows:} The function $f$ is defined as the composition of a discrete sequence of functions $g_1, \dots, g_n$, where the Jacobian determinant of $f$ can be efficiently computed by
\begin{equation}
    \det J_f = \det J_{g_1} \cdot \ldots \cdots \det J_{g_n}.
\end{equation}
One such approach is to define the $g_i$ as ``coupling layers'', which transform one subset of the variables conditioned on the complementary subset, resulting in a simple triangular Jacobian~\cite{dinh2014nice,dinh2016density}. Each $g_i$ is parametrized by free parameters which can collectively be optimized to reproduce a target distribution.

\paragraph{Continuous learnable flows:} The function $f$ is defined by integrating an ODE~\cite{Chen:2018wjc,zhang2018monge,zhang2021diffusion}. To ensure $f$ is a diffeomorphism, the ODE can be defined in terms of the gradient of a scalar function $\varphi$ as 
\begin{equation}
    \frac{d}{dt} U(t) = \nabla \varphi(U(t); t),
\end{equation}
giving $U(t) = V + \int_0^t dt' \, \nabla \varphi(U(t'); t') \big|_{U(0) = V}$ and $f[V] = U(T)$ in terms of the total flow time $T$.
For gauge fields, the differentiation, integration, and addition must be appropriately defined on the manifold of gauge configurations~\cite{Luscher:2009eq}. The Jacobian determinant can also be computed with an analogous ODE,
\begin{equation}
    \ln \det J = - \int_0^T dt \, \nabla^2 \varphi(U(t); t).
\end{equation}
Here, the scalar function $\varphi$ is parameterized by free parameters which can be optimized to reproduce a target distribution.
The trivializing map of Ref.~\cite{Luscher:2009eq} is a particular parameter-free choice of continuous flow.

Optimizing the parameters of a flow defined by one of the methods above requires a target ``loss function'' to be minimized, typically via stochastic gradient descent. A useful loss function for the lattice field theory context should provide a measure of the distance between the model distribution $q[U]$ and the known target distribution $p[U] = \exp(-S[U]) / Z$ without requiring many (or any) existing samples of the distribution. These features set apart the challenge of optimizing flow-based models for lattice field theory from many other machine learning contexts. Taking advantage of the explicitly known action $S[U]$, the reverse Kullback-Leibler divergence~\cite{kullback1951information} can be used to \emph{self-train} a model~\cite{zhang2018monge}, i.e., to optimize the flow function $f$ by performing stochastic gradient descent using only samples of field configurations drawn from the model and evaluations of the action on those configurations. This concept of self-training is similar in spirit to methods applied in non-flow approaches, such as for Self-Learning Monte Carlo methods~\cite{Huang:2017,Liu2017SLMC,Liu2016SLMCFermion,Xu:2017vug,Nagai2017SLMCCT,Shen2018SLMCDNN,Chen:2018aib,Liu:2019xnb,Nagai:2020ugq,Medvidovic:2020vum,Nagai:2021bhh,Nagai:2023fxt}.

\subsection{Application to scalar field theory}
\label{sec:csd-scalar-theory}
\begin{figure}
    \centering
    \begin{tikzpicture}
    \definecolor{beige}{RGB}{250,244,237};
    \node[fill=beige,rounded corners,inner sep=0pt] (coupling) at (0,0) {\includegraphics[height=4cm]{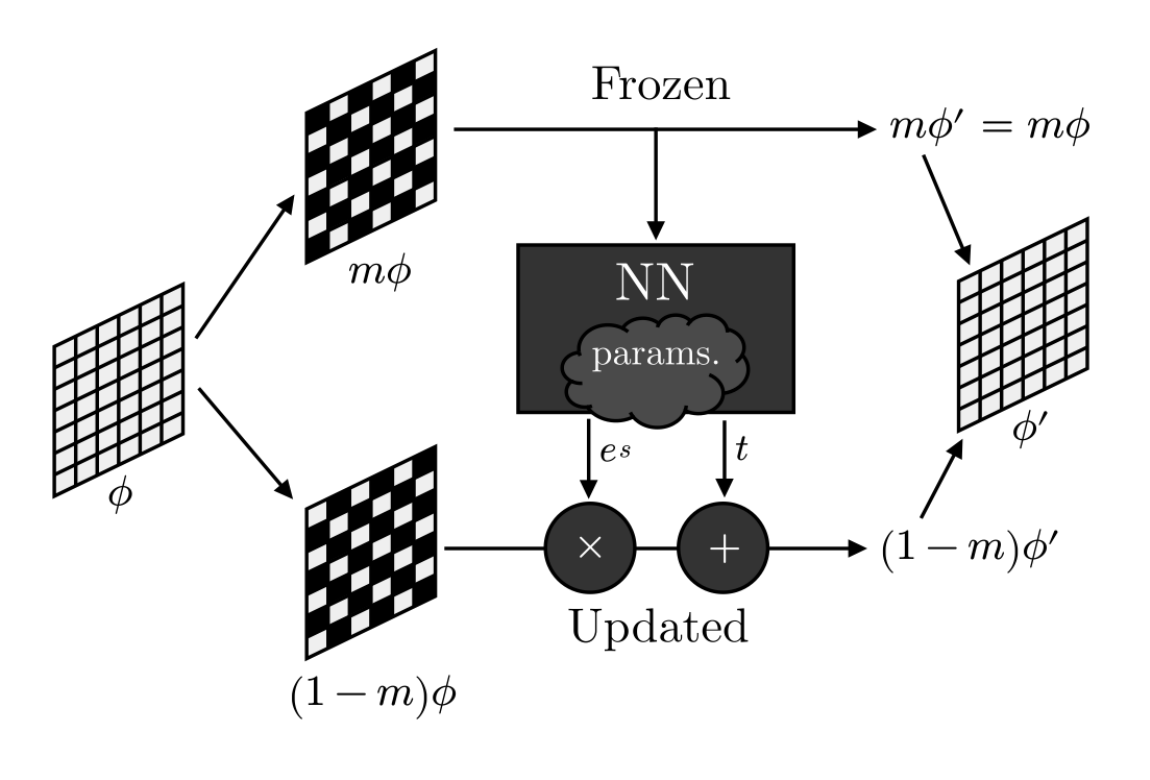}};
    \node (flow) at (7,0) {\includegraphics[width=7cm]{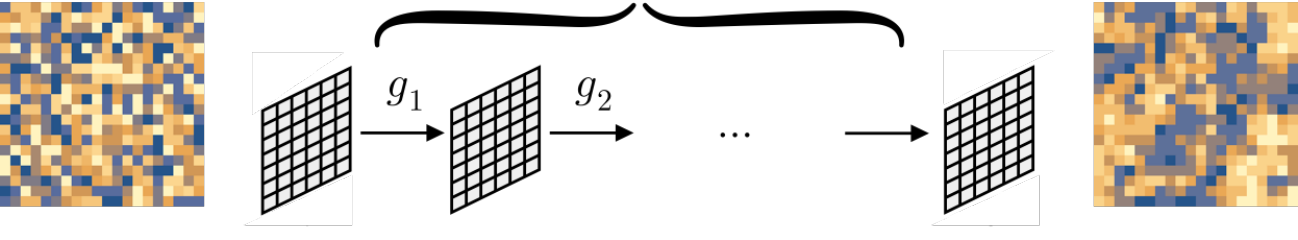}};
    \node[above=-.1cm of flow] {Flow $f$};
    \begin{pgfonlayer}{bg}
    \node[above right=-.77cm and -5.15cm of flow,fill=beige,circle,inner sep=5pt] (circle) {};
    \draw[beige,line width=1mm] (coupling.east)+(-1cm,1cm) to [bend left=45] (circle.center);
    \end{pgfonlayer}
    \end{tikzpicture}
    \caption{Discrete learnable flow architecture for scalar field theory demonstration in Ref.~\cite{Albergo:2019eim}.}
    \label{fig:scalar-flow}
\end{figure}
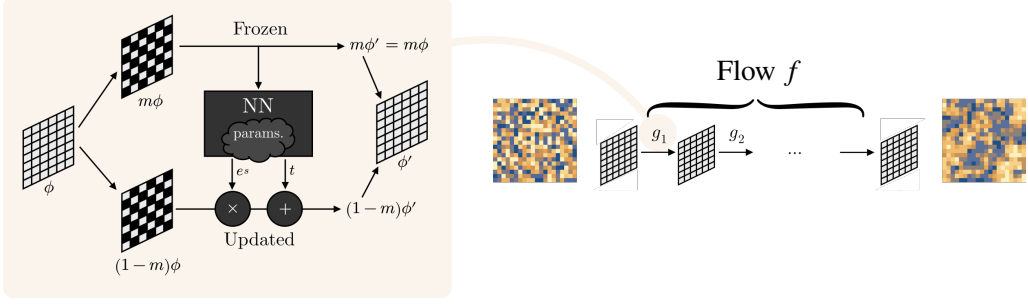

In Ref.~\cite{Albergo:2019eim}, these concepts were applied to demonstrate for the first time that critical slowing down could be essentially eliminated using flow-based models. The focus of this early study was an interacting scalar field theory on a $1+1$D lattice, described by the action
\begin{equation} \label{eq:scalar-action}
    S[\phi] = \sum_x \partial_\mu \phi(x) \partial^\mu \phi(x) + \frac{M^2}{2} \phi(x)^2 + \lambda \phi(x)^4.
\end{equation}
In this work, flow-based models used discrete learnable flows with a coupling layer architecture that alternately transformed even and odd sites conditioned on the complementary subset~\cite{dinh2014nice,dinh2016density}. Each such coupling layer locally applies the affine transformation $\phi'(x) = e^{s(x)} \phi(x) + t(x)$ with parameters $s(x)$ and $t(x)$ determined from neural network outputs, as shown on the left side of Fig.~\ref{fig:scalar-flow}. This transformation has a triangular Jacobian with simple determinant,
\begin{equation}
J_{ij} \equiv \partial \phi'_i / \partial \phi_j =
    \begin{bmatrix}
    I &  \\
    \blacksquare & \delta_{ij} e^{s_i}
    \end{bmatrix}
\quad \implies \quad
\ln \det J = \sum_i s_i .
\end{equation}
The right side of Fig.~\ref{fig:scalar-flow} schematically depicts the composed sequence of coupling layers which, when optimized, transform configurations sampled from a simple Gaussian prior distribution to configurations distributed approximately according to $p[\phi] = \exp(-S[\phi]) / Z$, in terms of the scalar field theory action $S[\phi]$ in Eq.~\eqref{eq:scalar-action}.

In the lattice field theory context, maintaining quantifiable uncertainties is also an important requirement. While a flow that only approximates a given target distribution produces biased samples, the use of a flow-based model with computable model density $q[\phi]$ allows any biases to be systematically corrected. For example, samples from the model can be used as proposals in a Markov Chain, with usual Metropolis acceptance steps applied to avoid bias~\cite{Albergo:2019eim}, or observables can be directly reweighted using the weights $p[\phi] / q[\phi]$~\cite{Nicoli:2019gun}.

By now, many works have demonstrated flow-based sampling for scalar field theory with larger lattice sizes, smaller lattice spacings, and in the broken symmetry phase~\cite{Hackett:2021idh,deHaan:2021erb,Nicoli:2020njz,Caselle:2022acb,Komijani:2023fzy,Gerdes:2022eve,Albandea:2023wgd,Albandea:2023Kl,Singha:2023cql}. Since this early work, there have also been many developments in the direction of incorporating gauge symmetries, incorporating dynamical fermions, and scaling to address challenging problems in lattice QCD, as we now discuss. A further overview of these developments can be found in Ref.~\cite{Cranmer:2023xbe}.

\subsection{Symmetries and application to gauge theory} \label{sec:symmetries-gauge-theories}
Symmetries of the lattice action play an important role in recovering the correct continuum limit, simplifying renormalization, and simplifying analysis. By applying Metropolis acceptance steps or reweighting, one is guaranteed to recover the correct physical distribution, including all symmetries. Nevertheless, incorporating symmetries directly in the definition of variational families of flow functions may improve both the training time and the final quality of the resulting flow-based model. Intuitively, all parameters of such a family describe the distribution in the meaningful physical directions, while symmetric directions remain ``flat'' by construction; this ensures that gradients provide good information about all parameters and may reduce the total number of model parameters required to construct models of similar quality.

The distribution of a flow-based model is defined by both the prior density $r[U]$ and the flow $f$. A sufficient condition to produce an invariant model is that, for all generators $t$ of the symmetry group,~\cite{kohler2020equivariant}
\begin{enumerate}
    \item The prior density is invariant, $r[t \cdot U] = r[U]$, and
    \item The flow is equivariant, $f[t \cdot U] = t \cdot f[U]$.
\end{enumerate}
The requirement on the prior distribution is often straightforward to satisfy for lattice field theory symmetries. Building flows that are equivariant under such symmetries requires more care.

\begin{figure}
    \centering
    \includegraphics[height=3cm]{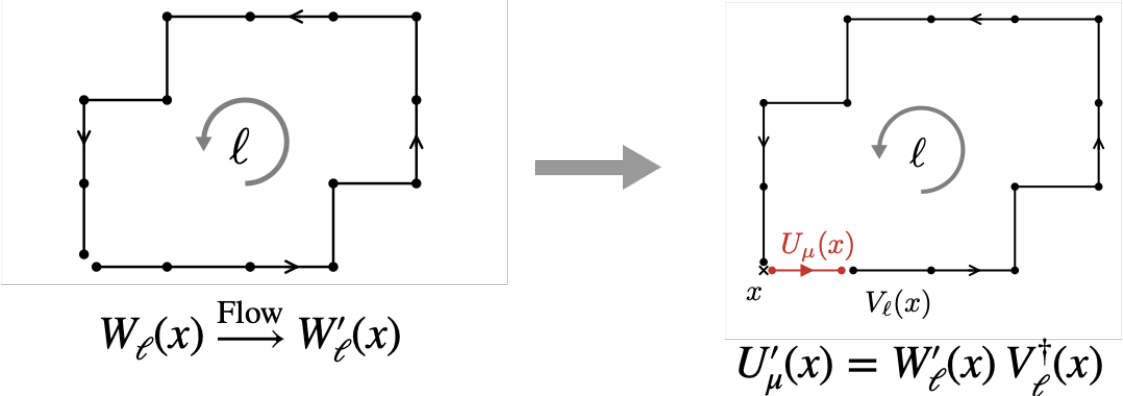}
    \caption{Gauge equivariant transformation defined in terms of inner transformations $W_\ell(x) \rightarrow W'_\ell(x)$ of untraced Wilson loops. Using the generalized staple $V_{\ell}(x)$, the transformation of each Wilson loop can be assigned to particular gauge links $U_\mu(x)$. Figure reprinted from Ref.~\cite{Kanwar:2021wzm}.}
    \label{fig:gauge-equiv}
\end{figure}

In practice, it is important to exactly incorporate gauge symmetry into flows to make progress in applying flow-based models to lattice gauge theories, such as QCD. A procedure to build gauge-equivariant coupling layers was introduced in Refs.~\cite{Kanwar:2020xzo,Boyda:2020hsi}, enabling computationally efficient discrete flows to be constructed for gauge theories. As shown schematically in Fig.~\ref{fig:gauge-equiv}, such coupling layers involve transforming gauge links $U_\mu(x)$ in terms of equivariant transformations of untraced Wilson loops $W_{\ell}(x)$, which satisfy the simpler same-site transformation $W_{\ell}(x) \rightarrow \Omega(x) W_{\ell}(x) \Omega^\dag(x)$ under gauge symmetry. In this case, custom neural networks suitable for acting on the compact $U(1)$ and $SU(N)$ manifolds also needed to be defined~\cite{Rezende:2020,Boyda:2020hsi}, highlighting the broader consideration that in many cases specialized machine learning developments need to be made to fit the lattice field theory context.

\begin{figure}
    \centering
    \begin{tikzpicture}
    \node[anchor=south east] at (0,0.3) {\includegraphics[width=8cm]{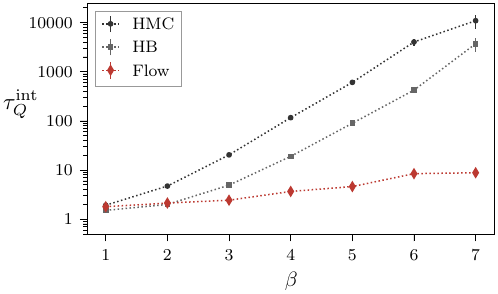}};
    \node[anchor=south west] at (0.4,2.8) {\includegraphics[width=4.9cm]{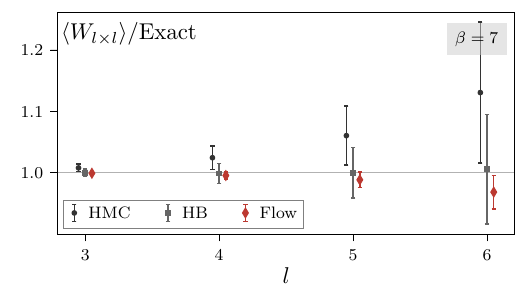}};
    \node[anchor=south west] at (0.4,0) {\includegraphics[width=4.5cm]{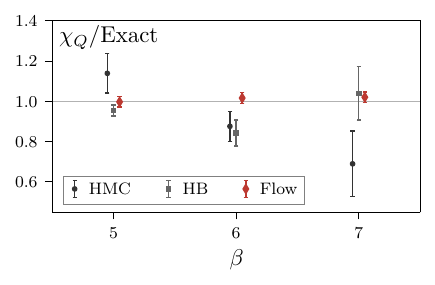}};
    \end{tikzpicture}
    
    \caption{Measured autocorrelation time of topological charge ($\tau^{\mathrm{int}}_Q$, left) and estimates of observables (Wilson loops $\left<W_{\ell \times \ell}\right>$ and topological susceptibility $\chi_Q$, right) in a $1+1$D $U(1)$ gauge theory~\cite{Kanwar:2020xzo}. Sampling with flow-based models (Flow) results in significantly reduced topological freezing in the continuum limit ($\beta \rightarrow \infty$) and more precise estimates versus Hybrid Monte Carlo (HMC) and Heatbath (HB) using identical ensemble sizes. Figures adapted from Refs.~\cite{Kanwar:2020xzo,Kanwar:2021wzm}.}
    \label{fig:u1-gauge}
\end{figure}

Figure~\ref{fig:u1-gauge} depicts an example of the significant improvements possible once such flows are constructed and optimized for lattice gauge theories, in this case for a $1+1$D $U(1)$ gauge theory~\cite{Kanwar:2020xzo}. Here, topological freezing in the continuum limit ($\beta \rightarrow \infty$) is significantly mitigated when sampling using flows. The result is far more precise observable estimates, especially for topological quantities such as the topological susceptibility, given the same ensemble sizes. In this example, the cost per configuration in the ensemble was roughly equivalent in all three methods.

The gauge-equivariant transformations described above are by no means the only approach to defining equivariant transformations. Recent works have developed and applied several new architectures for gauge-equivariant transformations that are suitable whenever invertibility is not required~\cite{Favoni:2020reg,Nagai:2021bhh,Namekawa:2022liz,Lehner:2023bba,Knüttel:2023w1,Aronsson:2023rli,Lehner:2023prf,Holland:2023lfx}. It is also possible to incorporate gauge equivariance into continuous flows by working with \emph{gauge-invariant} scalar potentials $\varphi$~\cite{Bacchio:2022vje}. In the view of the author, developing architectures for gauge-equivariant flow transformations that are both efficient and expressive remains one of the areas of this research requiring the most further investigation.

\subsection{Fermions and application to QCD}
\label{sec:fermions-qcd}
It is noteworthy that the methods described so far are already sufficient to apply normalizing flows to theories with fermions and theories in arbitrary spacetime dimensions, including the particularly challenging task of lattice QCD, as demonstrated by exploratory studies performed in Refs.~\cite{Albergo:2021bna,Albergo:2022qfi,Bacchio:2022vje,Boyle:2022xor,Abbott:2022hkm,Abbott:2023thq}. These studies have shown that there are no theoretical obstacles to flow-based sampling for almost all lattice field theories of interest. However, how to best design and optimize flow-based models in these more general contexts remains an area of active study, as we now discuss.

\section{Current research}

\subsection{New paradigms} \label{sec:new-paradigms}
Though motivated by the problem of critical slowing down, using flows to transform between lattice field theory distributions opens the door to several potential applications beyond directly sampling field configurations. These new paradigms are outlined below.

\begin{figure}
    \centering
    \includegraphics[width=6cm]{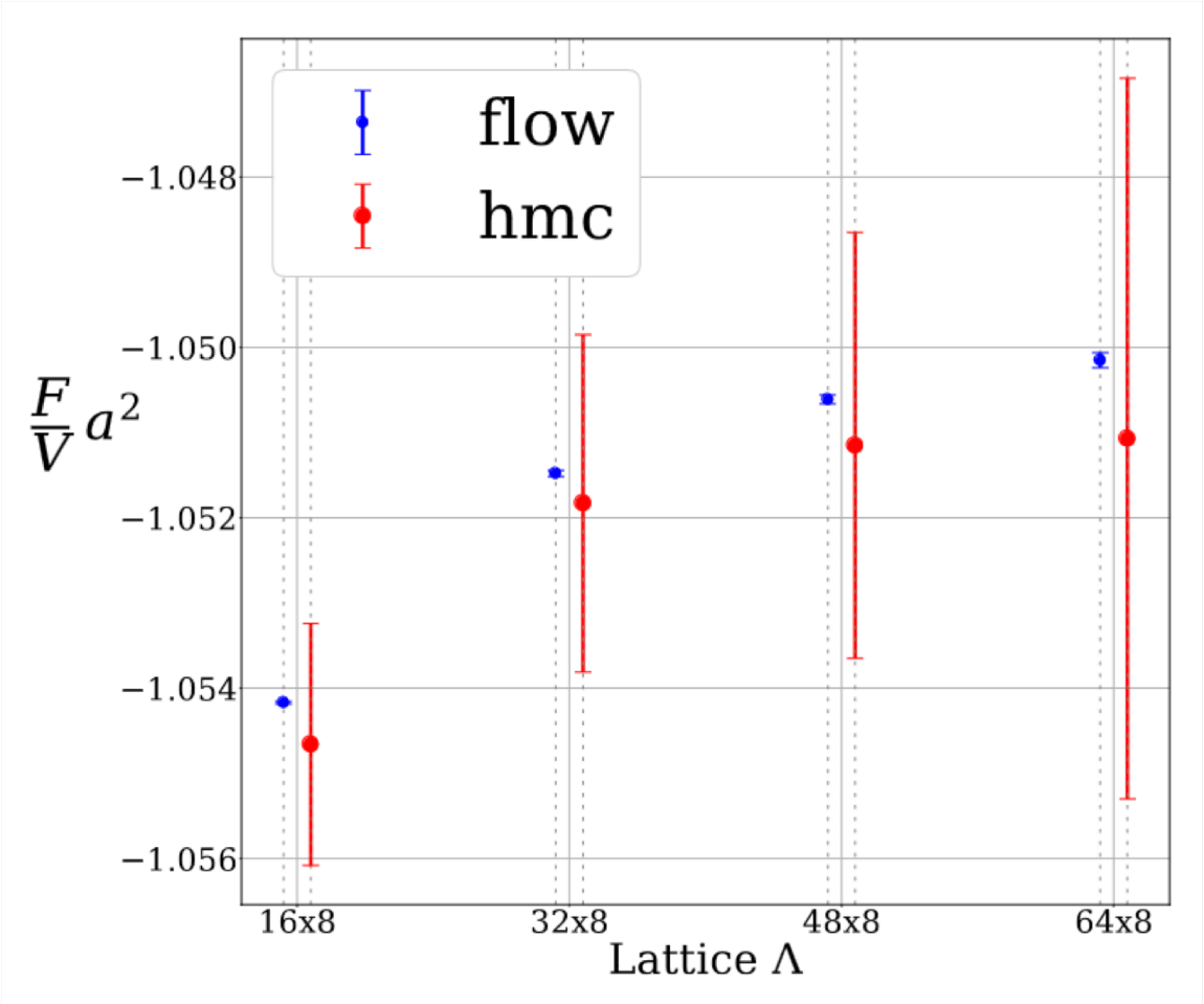}
    \caption{Free energy estimates in a scalar $\phi^4$ theory using a flow-based estimator (flow) vs a naive estimator using Hybrid Monte Carlo data (hmc). Figure reprinted from Ref.~\cite{Nicoli:2020njz}.}
    \label{fig:free-energy}
\end{figure}

\paragraph{Partition functions.}
The reverse Kullback-Leibler divergence used to optimize flows provides a lower bound on the partition function, $Z$, or conversely an upper bound on the free energy, $-\ln {Z}$~\cite{zhang2018monge}. 
Building on this observation, an unbiased, low-variance estimator for the partition function was recently introduced~\cite{Nicoli:2019gun,Nicoli:2020njz}. Figure~\ref{fig:free-energy} depicts the improvements in precision that was for example achieved in Ref.~\cite{Nicoli:2020njz} using this method to estimate the free energy of a scalar field theory. No modifications to the flow-based model architecture or training are required to apply this method, meaning that much more precise thermodynamic observables automatically become available if flow-based Monte Carlo sampling is used.

\paragraph{Parameter dependence.}
Flows can be defined with additional dependence on the bare physical parameters of the lattice action. In this case, if the parameterization of the flow is sufficiently flexible, one could hope to simultaneously define flows across an entire range of action parameters~\cite{dibak2022temperature,Gerdes:2022eve,Singha:2023cql}. Just as ordinary machine learning models have demonstrated the ability to effectively interpolate and extrapolate from a given data set, a similar training strategy could be applied for flows. Optimizing a parameterized family of flows on several representative choices of bare couplings, one aims to use the resulting model to efficiently interpolate and extrapolate to sampling a nearby values of the couplings. This for example could save significant computational effort when tuning bare parameters before ensemble generation.

\paragraph{Correlated samples.}
In addition to considering a family of flows from a common prior distribution to theories defined by many choices of physical parameters, as above, one can also construct flows that map \emph{between} theories at nearby points in parameter space~\cite{Abbott:2024xxx}. This approach has the potential to produce low-noise estimates of observables that require correlated samples from a pair of actions, for example in the Feynman-Hellmann method for evaluating matrix elements. Whereas flow-based direct sampling requires optimizing a flow between a simple prior distribution and more challenging target distribution, which typically are very different, this approach to correlated samples only requires constructing flows between very similar distributions.

\paragraph{Parallel tempering.}
Parallel tempering involves the execution of several Markov chains in parallel, each sampling an action interpolating between a distribution that is easier to sample and the target distribution~\cite{Swendsen:1986vqb}. The efficiency of the method is controlled by the exchange rate between neighboring chains. By using flows to perform an exchange, a high acceptance rate could be achieved, with a perfect flow corresponding to a $100\%$ acceptance rate~\cite{Invernizzi:2022,Abbott:2024yyy}. This approach is related to out-of-equilibrium simulations, which can similarly benefit from using normalizing flows as intermediate steps~\cite{Caselle:2022acb,Bonanno:2023ier}, and to the general framework of stochastic normalizing flows~\cite{wu2020stochastic}.

\paragraph{Sign problems.}
By defining flows to map into the complexified domain of field configurations, similar techniques may also be used to define and optimize deformations of the lattice path integral, which continue to be explored as potential solutions to sign problems~\cite{Lawrence:2021izu,Pawlowski:2022rdn,Rodekamp:2022xpf,Rodekamp:2023byu,Detmold:2023kjm}.

\subsection{Practical gains}
Using flow-based sampling in state-of-the-art lattice calculations would not only address difficult autocorrelations arising from critical slowing down, but would also practically simplify calculations in several ways, as outlined below.

\paragraph{Parallel sampling.}
In a Monte Carlo scheme where samples are drawn directly, many such samples can be produced in parallel. This structure maps very naturally onto the large multi-node machines presently used for state-of-the-art lattice QCD calculations, and results in almost perfect weak scaling. This benefit carries over even to the context of flow-based Markov Chain Monte Carlo, where normalizing flows are used to propose steps that are accepted or rejected using a Metropolis step. Despite the apparent serialization, the proposals are independent of the state of the chain, meaning they can be generated in parallel and arbitrarily in advance.

\paragraph{Storage-free ensembles.}
A surprising benefit of direct flow-based sampling is the possibility of eliminating long-term storage (and similarly communication) costs of ensembles of field configurations. With flow-based sampling, each configuration is produced independently starting from an easily-constructed sample of a prior distribution. In general, the sample from the prior distribution can be replaced with just the \emph{state of the random number generator} that was used to produce it, which is a much smaller object. For example, if each sample from the prior distribution is produced using a known seed, it would be sufficient to store or communicate the seed alone to be able to deterministically reconstruct the final field configuration on demand. This is a classic tradeoff between data size versus additional computation which is exploited in traditional compression algorithms. In this sense, a good normalizing flow can be considered an excellent compression scheme for lattice field configurations!

\subsection{Towards QCD at scale}
Among many interesting applications of lattice simulations, state-of-the-art lattice QCD represents by far the biggest computational challenge. At the same time, the field stands to make significant progress if a breakthrough in Monte Carlo sampling for lattice QCD could be achieved. As such, many developments in flow-based sampling have been oriented towards the goal of lattice QCD. By now, significant progress has been made in this direction.

This progress can be viewed in three broad stages:
\begin{itemize}
    \item Proof-of-principle studies of solving critical slowing down, incorporating gauge symmetries, and incorporating fermions (see Secs.~\ref{sec:csd-scalar-theory}, \ref{sec:symmetries-gauge-theories}, and \ref{sec:fermions-qcd});
    \item Works addressing architectural and training challenges presented by dynamical fermions; and
    \item Applications to higher spacetime dimensions and more degrees of freedom.
\end{itemize} 
The progress in the latter two stages is discussed briefly below. In the view of the author, we are now seeing progress in the final stage that will result in impact to some state-of-the-art calculations in the next couple of years, likely first through one or more of the new paradigms discussed in Sec.~\ref{sec:new-paradigms} rather than direct flow-based ensemble generation.

\begin{figure}
    \centering
    \includegraphics[width=8cm]{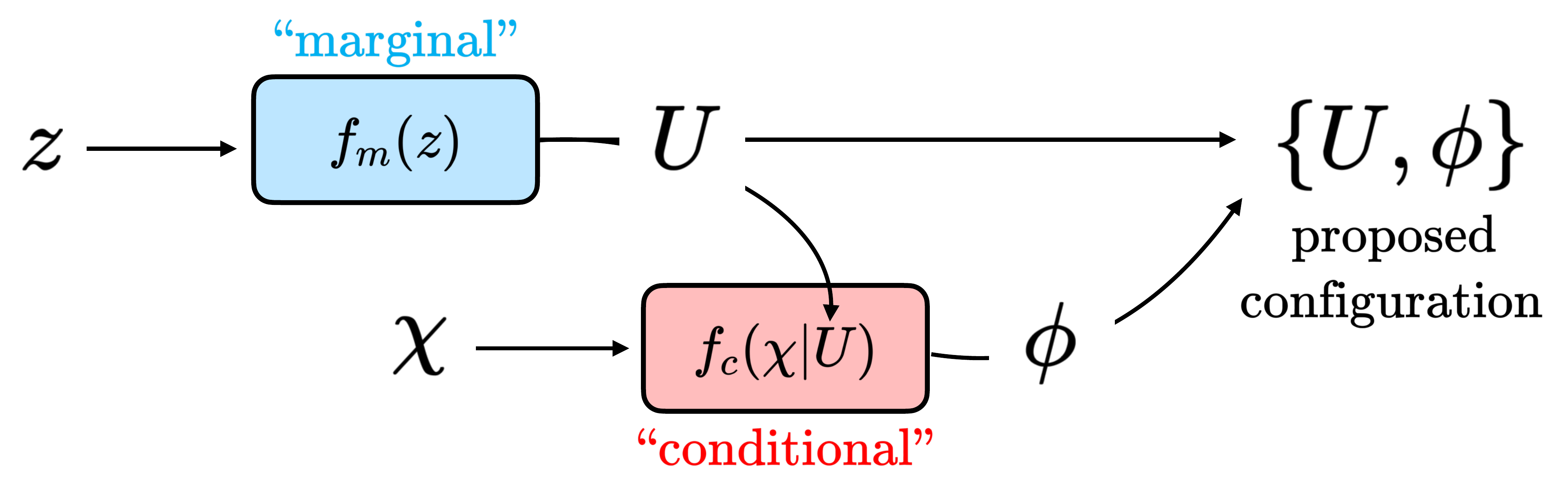}
    \caption{Structure of a normalizing flow model that jointly produces a bosonic field $U$ (for example a gauge field) and a pseudofermion field $\phi$. In this case, a marginal model first produces the bosonic field, then a conditional model produces the pseudofermion conditioned on the bosonic field. Figure reprinted from Ref.~\cite{Abbott:2022zhs}.}
    \label{fig:joint-model}
\end{figure}

\paragraph{Dynamical fermions.}
In lattice calculations, dynamical fermions are normally integrated out to produce an effective action over the remaining bosonic degrees of freedom that is written in terms of a determinant of the Dirac operator. In principle, flow-based models can be optimized to reproduce this target distribution just as for an ordinary bosonic theory. Early studies demonstrated that it was possible to train models to closely reproduce this effective action using exact evaluations of the determinant~\cite{Albergo:2021bna,Albergo:2022qfi}. However, just as for Hybrid Monte Carlo (HMC), this approach does not scale favorably with the number of lattice sites. This challenge can be avoided as in HMC by using the pseudofermion representation of the determinant~\cite{Albergo:2021bna,Abbott:2022zhs,Abbott:2022hkm}, for example by using a joint model as shown in Fig.~\ref{fig:joint-model} to sample bosonic fields and pseudofermions simultaneously. The ``marginal'' normalizing flow used to sample the effective action is combined with a ``conditional'' flow that produces pseudofermion samples along with a \emph{normalized} estimate of their probability density, effectively providing an estimator of the determinant. In addition, Ref.~\cite{Abbott:2022zhs} showed that the method can be combined with even-odd or Hasenbusch preconditioning as usual, and noise from the pseudofermion estimator of the determinant can be reduced by averaging over multiple samples.

\paragraph{Higher dimensions and more degrees of freedom.}
While the technology described above is sufficient to apply flow-based sampling to almost any lattice field theory of interest, there are practical challenges associated with training models with many degrees of freedom. There is also still much ground that can be gained by experimenting with varying the architectures of flow-based models, as has been demonstrated by recent advances~\cite{Finkenrath:2022ogg,Abbott:2023thq,Bacchio:2022vje}. With such refinements and more dedicated studies, flow-based sampling has now been applied to the $N_f = 2$ Schwinger model with up to $128 \times 128$ lattice sites~\cite{Finkenrath:2022ogg}, to $3+1$D Yang-Mills~\cite{Abbott:2023thq,Bacchio:2022vje}, and to $N_f = 2$ QCD in a small volume~\cite{Abbott:2022hkm}.

\subsection{Frontiers of development}
To take the remaining steps towards flow-based sampling for lattice field theories at state-of-the-art scale, including towards improved lattice QCD calculations, continued development is needed. The many frontiers of these developments can unfortunately not be covered here in detail. Instead, only the most recent developments in several promising directions are discussed below.

\paragraph{New ways to construct flows.}
The flexibility of flows composed from a discrete sequence of coupling layers depends on each transformation being sufficiently expressive and their composition enabling further expressivity. As discussed above, the requirement of enforcing gauge equivariance in coupling layers restricts their form and likely their expressivity. Recent work has sought new ways to define gauge equivariant layers that are both individually expressive and compose well~\cite{Abbott:2023thq}, enabling applications to theories in higher spacetime dimensions.

Recently, several works have also investigated learnable continuous flows defined by the integration of a continuous transformation~
\cite{Boyle:2022xor,Bacchio:2022vje}. Using a physics-inspired basis for the learnable potential, parameterized by collections of Wilson loops of varying sizes, promising results have been demonstrated for sampling gauge theories using very few free parameters. The challenge, as for discrete flows, is to parameterize flows that can be expressive enough to capture the difficult distributions associated with realistic lattice gauge theories in higher spacetime dimensions and at fine lattice spacings. Additionally, for continuous flows, accurate integration of the ODEs defining the flow and its Jacobian are required. Nevertheless, the simplicity of incorporating symmetries in this approach is attractive---defining families of \emph{invariant} potentials tends to be a less complex task than defining \emph{equivariant} coupling layers.

It is noteworthy that the benefits of both approaches may be combined: Ref.~\cite{Abbott:2023thq} introduced \emph{residual coupling layers}, which define discrete invertible transformations by the gradient of a scalar potential. This architecture is currently being further explored.

Meanwhile, recent works have explored improving normalizing flows by incorporating stages that work in a Fourier momentum basis, rather than the lattice position basis~\cite{Chen:2022ytr,Komijani:2023fzy}. These efforts have so far demonstrated promising results for quantum mechanics problems and scalar field theory.

rFinally, normalizing flows constructed from a hierarchical sequence of transformations is another natural approach to model the hierarchy of scales in typical lattice field theory problems. This can be considered an application of ``in-painting'' which has been studied thoroughly in the image generation community, including through the use of normalizing flows~\cite{dinh2014nice}. This approach has been explored in only a few lattice field theory examples so far~\cite{Matsumoto:2023akw,Abbott:2024zzz}.

\paragraph{New ways to incorporate flows.}
The focus of this review has largely been on applying normalizing flows to convert easily produced samples of a simple prior distribution to independent samples of a more complicated distribution. However, flows as flexible invertible transformations of field configurations can potentially be incorporated in lattice field theory calculations in many ways. Some of these ways have already been discussed in the new paradigms covered in Sec.~\ref{sec:new-paradigms}:
\begin{itemize}
    \item Normalizing flows may be used to produce a correlated field configuration distributed according to one action from a field configuration drawn from the other.
    \item Similarly, flows may be used to map a configuration to a more typical configuration under a target action when proposing exchanges between neighboring chains in parallel tempering.
    \item Flows may be used to define contour deformations to solve sign problems.
\end{itemize}

Flows can also be incorporated into more standard Markov Chain sampling approaches. On one hand, both the trivializing map and more general flows have been applied to transform to a more easily sampled distribution on which standard HMC methods can be applied for faster mixing~\cite{Luscher:2009eq,Engel:2011re,Boyle:2022xor,Albandea:2023wgd,Albandea:2023Kl,Matsumoto:2023akw}. On the other hand, several recent works have explored \emph{flow-based HMC}, in which HMC trajectories are themselves defined by learned flow transformations~\cite{Foreman:2021ljl,Foreman:2023ymy}.
In contrast to directly producing samples from a hierarchical normalizing flow, as discussed above, recent works have also considered using domain decomposition to combined flows in inner blocks with traditional Markov Chain steps in an outer procedure~\cite{Finkenrath:2022ogg,Komijani:2023fzy,Faraz:2023xdi}. This approach has the benefit of exploiting the expected decorrelation of subvolumes of field configurations that are separated by sufficient physical distance, circumventing the known challenges surrounding directly sampling field configurations at large physical volume~\cite{Abbott:2022zhs}.

\paragraph{Application to new theories.}
In recent years, several other lattice theories have been approached using normalizing flows. In particular, flows have been applied to the effective Nambu-Goto action of a string worldsheet~\cite{Caselle:2023mvh,Caselle:2023ie},
the XY model~\cite{dibak2022temperature,Faraz:2023xdi}, and $CP^{N-1}$ models~\cite{Bonanno:2023ier,Chamness:2023/i}. In the view of the author, it is particularly interesting to further study whether critical slowing down can be eliminated for $CP^{N-1}$ models, as these models reproduce many of the features of QCD~\cite{DAdda:1978vbw,Witten:1978bc} while remaining computationally inexpensive. Parallel tempering methods, potentially combined with normalizing flows, have made progress against topological freezing in $CP^{N-1}$ models~\cite{Hasenbusch:2017unr,Berni:2019bch,Bonanno:2023ier}, and it remains to be seen whether all significant critical slowing down effects can be addressed by such means or other flow-based sampling approaches.

\paragraph{New software.}
Exploring normalizing flow techniques requires implementing models with components and architectures not currently used in the machine learning community, precluding the use of off-the-shelf implementations. Some early steps have been taken in the direction of making code specific to lattice field theory available in the form of a Jupyter notebook~\cite{Albergo:2021vyo} and a Julia package~\cite{Tomiya:2022meu}, each with implementations of specific flow models, as well as a python implementation of flow-based HMC~\cite{Foreman:2021ljl}. Recently, Ref.~\cite{Nicoli:2023JO} introduced a fully general package aimed at collecting components and tools across the various applications of flows for lattice field theory. In addition, one of the main practical challenges of studying normalizing flows at scale is the absence of highly optimized codes such as those available for traditional HMC simulations. While machine learning libraries provide the ability to target hardware accelerators and perform relatively optimized operations, reaching state-of-the-art scales will require significant further development, especially towards efficient performance in a distributed computing context.

\begin{figure}
    \centering
    \begin{minipage}{\textwidth}
    \raisebox{-0.5\height}{\includegraphics[width=6.5cm]{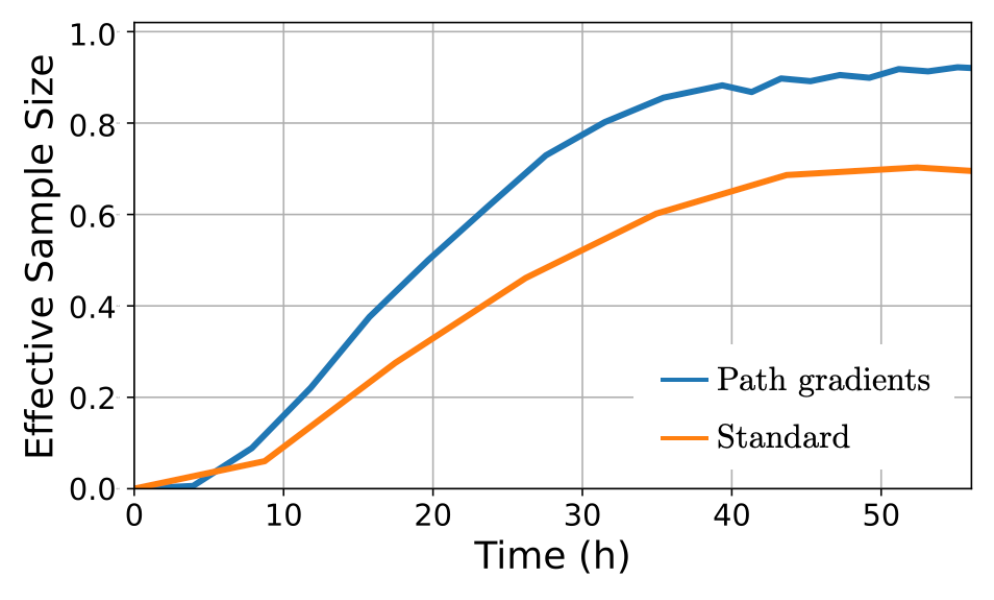}}
    \hspace{1cm}
    \raisebox{-0.5\height}{\includegraphics[width=7cm]{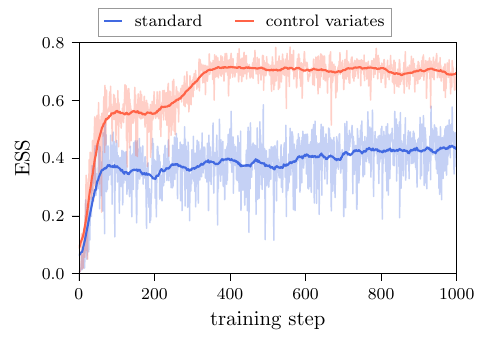}}
    \end{minipage}
    \caption{Left: Model quality (effective sample size) as a function of optimization time using the standard gradient estimator versus an improved path gradient estimator for a $1+1$D scalar field theory. Right: Model quality (effective sample size, ESS) as a function of the number of gradient descent steps using the standard gradient estimator versus an improved gradient estimator using control variates for a $3+1$D $SU(3)$ gauge theory. Figures adapted from Refs.~\cite{vaitl2022path, Abbott:2023thq}.}
    \label{fig:control-variates}
\end{figure}

\paragraph{New training methods.}
The choice of training (optimization) scheme used to determine a good flow function from a variational family can have significant effects both on the data efficiency of training and the final quality of the resulting flow-based model. Early works largely used simple stochastic estimators of gradients of the reverse Kullback-Leibler divergence for stochastic gradient descent optimization. Recently, however, a number of works have explored lower-variance estimators of the same gradients as a means of improving the convergence, demonstrating in several cases significantly better training times and results~\cite{Bialas:2022iro,vaitl2022path,vaitl2022gradients,Bacchio:2022vje,Abbott:2023thq,Bialas:2023fyj}. For example, Fig.~\ref{fig:control-variates} demonstrates the stark improvement in model quality when working with improved estimators to optimize flow-based models for examples of $1+1$D scalar field theory~\cite{vaitl2022path} and $3+1$D $SU(3)$ gauge theory~\cite{Abbott:2023thq}.

For theories with a spontaneously broken symmetry, the target distribution features either multiple isolated modes, corresponding to distinct classical minima of the action, or extended modes, corresponding to a continuum of minima. These structures can also present difficulties to optimization using the reverse Kullback-Leibler divergence. Recently, several works have studied these challenges, proposing several practical methods to avoid training difficulties~\cite{Hackett:2021idh,Nicoli:2023qsl}.

Finally, further exploration has demonstrated that the time for optimization and the final model quality are both sensitive to many hyperparameter choices of the models and optimization procedure~\cite{Abbott:2022zsh}. It is clear that in many cases training costs can be improved, sometimes by orders of magnitude. One systematic study was performed of the scaling of training costs when working with a particular model architecture, demonstrating a steep increase in the required training time as the continuum limit was approached~~\cite{DelDebbio:2021qwf}. This work highlights the need to continue to investigate the training procedure and model architecture. Whether similar scaling holds for more modern flow-based models at large scales remains an interesting open question, meriting careful future investigation.

\section{Outlook}

In the past few years, rapid progress has been made towards the goal of implementing effective normalizing flows for lattice field theory configurations. Flow-based models have been shown to essentially eliminate critical slowing down and topological freezing for certain theories. New components and architectures have been developed to extend flow-based models to apply to a wide variety of theories: theories with gauge and fermionic degrees of freedom, theories in arbitrary spacetime dimensions, and theories with non-trivial phase structure. Significant progress has specifically been made in the direction of applying flow-based sampling for the particularly challenging task of large-scale lattice QCD calculations, including several applications to theories with light dynamical fermions, pure-gauge theories in $3+1$D, and $N_f = 2$ QCD in a small volume.

There are of course many open questions remaining and exciting new directions of research currently under way. For example, recent works on continuous normalizing flows have demonstrated that, for several gauge theory applications, a particularly parameter-efficient representation of normalizing flows is possible, inspired by a generalization of trivializing maps. Whether and how these results can be combined with existing work on discrete normalizing flows remains to be seen. Meanwhile significant strides have been made in our understanding of how to optimize normalizing flows for lattice field theory. Yet how to best architect and optimize normalizing flows to minimize training costs, and the precise scaling of those costs, is still unclear. Developing a mature and optimized software library on which the community can build particular models and ensembles is also an important step towards applying flow-based models at state-of-the-art scales. We have seen recent, ongoing efforts in this direction.

It is likely that we will see the first demonstration of practical benefits beyond a proof of principle in applications that incorporate normalizing flows as components within existing calculations, for example by reducing variance in contexts where correlated samples are required or perhaps in improvements to HMC through the inclusion of normalizing flow components. The long term future of such a young research area cannot be reliably predicted. However, given our present understanding of the challenges and features of normalizing flows, we can consider one potential avenue: It is possible that the maximum impact on Monte Carlo sampling for lattice field theory will be achieved through a hybrid method, taking advantage of the natural hierarchy of scales of most lattice field theory problems to organize Monte Carlo sampling in a nested fashion, with independent sampling of inner subvolumes via normalizing flows and traditional Markov Chain updates of boundary degrees of freedom. Several works are beginning to investigate this approach already. This scheme might naturally be combined with practical improvements, such as highly parallelized evaluation across subvolumes, elimination of storage costs of inner degrees of freedom, and exponential variance reduction techniques based on hierarchical sampling~\cite{Luscher:2001up,Ce:2016idq}. If such a sampling scheme can be realized, it has the potential to drastically increasing the precision of lattice field theory calculations across a variety of fields, enabling a deeper understanding of physics ranging from condensed matter theory to QCD and particle physics beyond the Standard Model.

\acknowledgments

I am grateful to the organizers for the invitation to present this review and outlook. I would also like to thank Phiala Shanahan for helpful comments on this manuscript and all of my collaborators for their significant contributions towards the portion of the work reviewed here that is our own. This work is supported by the Swiss National Science Foundation (SNSF) under grant 200020\_200424.

\bibliographystyle{JHEP}
\bibliography{main.bib,self.bib}

\end{document}